\documentclass[aps,prb,twocolumn,superscriptaddress,groupedadress,floatfix,longbibliography]{revtex4-2}

\usepackage{amsmath,amssymb} 
\usepackage{physics}
\usepackage{bm} 
\usepackage{graphicx} 
\usepackage{comment} 
\usepackage{textcomp} 

\usepackage{enumitem}
\setlist{noitemsep,leftmargin=*,topsep=0pt,parsep=0pt}

\usepackage{xcolor} 
\definecolor{lightgray}{gray}{0.6}
\definecolor{medgray}{gray}{0.4}

\usepackage{hyperref}
\hypersetup{
colorlinks=true,
urlcolor= blue,
citecolor=blue,
linkcolor= blue,
}

\begin{document}

\title{Thermal Evolution of Skyrmions in Synthetic Ferrimagnets of Co/Gd Heterostructure for Topological Spintronic  Applications}

\author{Bhuvneshwari Sharma}
\affiliation{Laboratory for Nanomagnetism and Magnetic Materials (LNMM), School of Physical Sciences, National Institute of Science Education and Research (NISER), An OCC of Homi Bhabha National Institute (HBNI), Jatni 752050, Odisha, India}

\author{Soumyaranjan Dash}
\affiliation{Department of Physical Sciences, Indian Institute of Science Education and Research (IISER) Mohali, Sector 81, S.A.S. Nagar, Manauli PO 140306, India}

\author{Shaktiranjan Mohanty}
\affiliation{Laboratory for Nanomagnetism and Magnetic Materials (LNMM), School of Physical Sciences, National Institute of Science Education and Research (NISER), An OCC of Homi Bhabha National Institute (HBNI), Jatni 752050, Odisha, India}

\author{Brindaban Ojha}
\affiliation{Laboratory for Nanomagnetism and Magnetic Materials (LNMM), School of Physical Sciences, National Institute of Science Education and Research (NISER), An OCC of Homi Bhabha National Institute (HBNI), Jatni 752050, Odisha, India}

\author{Debi Rianto}
\affiliation{Department of Physics, Durham University, Durham DH1 3LE, U.K.}

\author{Del Atkinson}
\affiliation{Department of Physics, Durham University, Durham DH1 3LE, U.K.}

\author{Sanjeev Kumar}
\email{sanjeev@iisermohali.ac.in}
\affiliation{Department of Physical Sciences, Indian Institute of Science Education and Research (IISER) Mohali, Sector 81, S.A.S. Nagar, Manauli PO 140306, India}

\author{Subhankar Bedanta}
 \email{sbedanta@niser.ac.in}
\affiliation{Laboratory for Nanomagnetism and Magnetic Materials (LNMM), School of Physical Sciences, National Institute of Science Education and Research (NISER), An OCC of Homi Bhabha National Institute (HBNI), Jatni 752050, Odisha, India}

\begin{abstract}
Synthetic ferrimagnetic (SFiM) multilayers offer a versatile platform for hosting skyrmions with tunable magnetic properties, combining the advantages of ferromagnets and antiferromagnets. Unlike synthetic antiferromagnets, SFiMs retain a finite magnetization that allows direct observation of magnetic textures while still benefiting from reduced dipolar fields and a suppressed skyrmion Hall effect. However, a systematic investigation of their temperature and field dependent magnetization evolution, including the labyrinthine-to-skyrmion transition in Co/Gd-based SFiMs, remains less explored. Here, we demonstrate the stabilization of ~70 nm-radius skyrmions at room temperature and reveal how the Co and Gd sublattices influence the temperature-dependent net magnetization. Further, we develop a microscopic spin model for SFiM incorporating the relevant magnetic interactions, which reproduces the experimental observations and captures the temperature-dependent magnetic phase evolution. This framework highlights the interplay of fundamental interactions controlling skyrmion stability in SFiM and provides a pathway for engineering heterostructures for topological spintronic applications.
                  
\end{abstract}
\maketitle
\section{Introduction}
Nanoscale spin textures with topological protection, known as skyrmions, have attracted considerable attention due to their stability, small size, and efficient current-driven motion \cite{Bogdanov1989, Rosler2006, Nagaosa2013, Fert2013, Ojha2025}. Their unique properties make them promising candidates for a variety of spintronic applications \cite{Everschor2018}. Skyrmions can serve as information carriers in high-density memory devices, where the presence or absence of a skyrmion encodes binary information \cite{Tomasello2014, Ding2015}. Beyond memory, skyrmions have also been explored as building blocks for unconventional computing devices, including logic gates, reservoir computing architectures, and neuromorphic systems \cite{Sisodia2022, Luo2018, Chauwin2019, Beneke2024, Dash2023}. Furthermore, their topological protection and solitonic behavior offer advantages for energy-efficient manipulation and robust operation under current-induced driving forces \cite{Fert2017, Sampaio2013, Ojha2023}. Despite their potential, several challenges limit the practical implementation of skyrmions. Skyrmion motion is often affected by pinning from defects, inhomogeneities, or grain boundaries, preventing ideal flow \cite{Litzius2017, Gruber2022, Gong2020}. Furthermore, the skyrmion Hall effect (SkHE), arising from their topological charge, can deflect skyrmions from intended paths, causing instability in device operation \cite{Jiang2017, Chen2017}. In addition, for applications such as magnetic data storage, thermal stability must be effectively controlled \cite{809134}. Recently, materials with reduced or compensated magnetization, such as synthetic antiferromagnets (SAFs) and ferrimagnets, have been explored to overcome some of these limitations \cite{Dohi2019, Pham2024, Woo2018, Mohanty2024}. SAFs, with two ferromagnetic layers coupled antiferromagnetically by a spacer layer, offer nearly zero net magnetization and suppress the SkHE. However, their vanishing net moment makes magnetic states difficult to detect using conventional magneto-optical techniques. \\
On the contary, synthetic ferrimagnets(SFiM), composed of two antiferromagnetically coupled sublattices, reduce the net magnetization and suppress the SkHE while retaining a small but finite moment, allowing skyrmions to be detected and driven at high speed  motion \cite{Woo2018, Ma2019, Kim2022}. Building on these concepts, SFiMs composed of transition metals (Co, Fe) and rare-earths (Gd, Tb) are governed by an indirect negative exchange, which aligns the transition-metal and rare-earth spins antiparallel to each other \cite{Nesbitt1962}. This negative exchange interaction stabilizes compact skyrmions by reducing the net magnetization, ensures their detectability, suppresses the SkHE by maintaining a small but nonzero magnetic moment, and thereby enables efficient current-driven manipulation \cite{Finley2016, Mishra2017, Siddiqui2018, Mallick2024}. SFiMs can be fabricated either as alloys or as multilayer heterostructures, offering versatile platforms for skyrmion-based spintronic devices. Recent studies on ferrimagnetic alloys highlight their promise for high-speed spintronic applications. Subsequent works have demonstrated current-driven domain-wall and skyrmion motion, zero-field skyrmion nucleation, and reduced Hall angles in CoGd alloys \cite{Kim2017, Caretta2018, Berges2022}. Composition tuning, such as varying Tb content, further enables control over skyrmion size and stability \cite{Xu2023}. While these alloy systems reveal the promise of ferrimagnetism for skyrmion control, SFiM multilayers offer additional flexibility through interface engineering and layer design. These systems have recently been explored for novel functionalities: for example, multilayers that integrate all-optical switching with current-driven domain dynamics highlighting their potential for opto-spintronic applications, as well as skyrmion-enhanced strain-mediated reservoir computing with high recognition and predictive performance.  \cite{Li2023, Li2025, Sun2023}.\\
However, the temperature dependent behavior and skyrmion stability has not been significantly studied. There are only a few studies that have addressed the role of interfacial physics and temperature-dependent effects in Co/Gd multilayers, and a comprehensive understanding of the mechanisms governing skyrmion stability remains lacking.0 While AFM-coupled skyrmion bubbles in Co/Gd multilayers have shown topological spin memory effects, where the skyrmion configuration is recovered after thermal cycling owing to their topologically protected spin structure \cite{Wang2022}. Other studies have focused solely on induced magnetic moments and flipped spin states at Co/Gd interface arising from proximity effects, without directly addressing skyrmion formation \cite{Brandao2025}. These micromagnetic and element-resolved investigations hint at the importance of interfacial interactions and Dzyaloshinskii–Moriya interactions (DMI), but a microscopic framework to describe the temperature-dependent evolution from labyrinthine domains into nanoscale skyrmions in SFiMs has not been established.\\
Here, we study Pt/Co/Gd multilayers to build a comprehensive understanding of temperature-dependent magnetization, skyrmion formation, and their robustness in SFiM systems, using a combination of bulk and localized experimental characterization together with theoretical modeling. The multilayer exhibits anisotropy near to spin reorientation transition (SRT) and antiferromagnetic coupling between Co and Gd sublayers, forming a ferrimagnetic configuration. Magnetic force microscopy (MFM) reveals labyrinthine domain patterns are transformed into isolated skyrmions under applied out-of-plane fields. Temperature-dependent studies show that the Co and Gd sub-lattice play a role in modulating the net magnetization. To gain microscopic insight into these effects, we employ a theoretical model that reproduces the observed magnetic textures and captures the temperature-dependent behavior, providing a systematic framework to explore how magnetic and thermal parameters influence skyrmion formation and stability in SFiMs.

\section{Methods}

\subsection{Sample growth and structural characterization}

Multilayer thin films with the structures of sample FS:
Ta(3)/Pt(5)/[Pt(2)/Co(1.5)/Gd(0.8)]$_5$/Ta(4)/Pt(2),
sample FR$_{1}$: Ta(3)/Pt(5)/[Pt(2)/Co(1.5)/Gd(0.8)]/Pt(4)
/Ta(2), and sample FR$_{2}$: Ta(3)/Pt(5)/[Pt(2)/Co(1.5)/
Gd(0.8)]/Ta(4)/Pt(2)
were grown on commercial silicon (Si) substrates with a 100 nm thick layer of thermally grown silicon dioxide (SiO$_2$).
 The numbers in parentheses indicate the thickness of each layer in nano-meters(nm). The deposition was done using a high-vacuum sputtering system (Mantis Deposition Ltd.) with a base pressure of $6\times 10^{-8}$ mbar. Tantalum (Ta), Platinum (Pt), and Cobalt (Co) layers were deposited via DC magnetron sputtering, while the Gadolinium (Gd) layer was deposited using RF magnetron sputtering. The initial Ta seed layer was critical for promoting the growth of a (111)-oriented Pt layer \cite{Parakkat2016}. This specific crystalline orientation of Pt is essential for inducing perpendicular magnetic anisotropy (PMA) in the subsequent Co and Gd layers, as it facilitates a strong interfacial magnetic anisotropy at the Pt/Co, and Co/Gd interfaces. A final protective Pt capping layer was deposited to prevent oxidation of the underlying magnetic layers. The structural parameters of the multilayer samples were quantified using X-ray reflectivity (XRR) measurements performed on a D8 Bruker Reflectometer, and the data was analyzed by fitting the experimental patterns using the GenX software package \cite{Bjorck2007}.

\subsection{Magnetic characterization and imaging}
The temperature-dependent magnetization was measured using Superconducting Quantum Interference Device (SQUID) magnetometer manufactured by Quantum Design, USA. Hysteresis loops were obtained at selected temperatures by sweeping the external magnetic field up to ±3T in both out-of-plane (OOP) and in-plane (IP) geometries, allowing evaluation of the magnetic anisotropy and coercivity as a function of temperature. Magnetic domain imaging of the multilayers was performed at room temperature via MFM manufactured by Attocube. To minimize probe–sample interaction, low-moment tips were used. MFM images were first acquired in the demagnetized state, followed imaging during the application of an increasing OOP magnetic field to study the field-induced evolution of the magnetic domain structure.

\subsection{Microscopic model and simulation}
In order to arrive at a minimal model for the generalized SFiM heterostructures (SFiMH), we note that the itinerant character of magnetism, the spin-orbit coupling (SOC), and the inter-layer interactions are the essential features of the magnetic system. The minimal set up that allows for inclusion of all these features is a two-layer model. The presence of a larger number of layers may help in stabilizing the magnetic order at higher temperature by avoiding the Mermin-Wagner theorem. In the model, this can be achieved by introducing a uniaxial anisotropy which is also known to be present in the SFiM. Therefore, we consider a bilayer triangular lattice, representing $l=1$ to the Co layer and $l=2$ to the Gd layer. The influence of Pt is included in the model via the presence of a Hund's rule coupling and SOC. The resulting Hamiltonian is given by,

\begin{eqnarray}
	H & = & \sum_{l=1}^2 H^{(l)} + H_{\rm int} \label{Ham}
\end{eqnarray}

Where,

\begin{align}
	H^{(l)} &= 
	- t^{(l)} \sum_{i,\gamma,\sigma} \left(c^\dagger_{i, l, \sigma} c_{i+\gamma,l, \sigma} + \text{H.c.}\right)-J^{(l)}_\text{H} \sum_{i} {\bf S}_i^{(l)} \cdot {\bf s}_i^{(l)} \nonumber \\
	&\quad
    - {\textrm i}\lambda^{(l)} \sum_{i, \gamma, \sigma \sigma'} c_{i, l, \sigma}^{\dagger} [\pmb{\tau}\cdot(\hat{\pmb{\gamma}} \times \hat{\bf{z}})]_{\sigma \sigma'} c_{i+\gamma, l, \sigma'} \nonumber \\
	&\quad 
- \sum_{i} \left[h_z S_i^{(l)z} + A_u^{(l)} \left(S_i^{(l)z}\right)^2 \right]  \label{Ham1}
\end{align}

\begin{align}
	H_{\rm int} &= J_{\rm AF} \sum_{i} {\bf S}_{i}^{(1)} \cdot {\bf S}_{i}^{(2)} \label{Ham2}
\end{align}

\begin{figure*}
\includegraphics[width=1.1 \columnwidth,angle=0,clip=true]{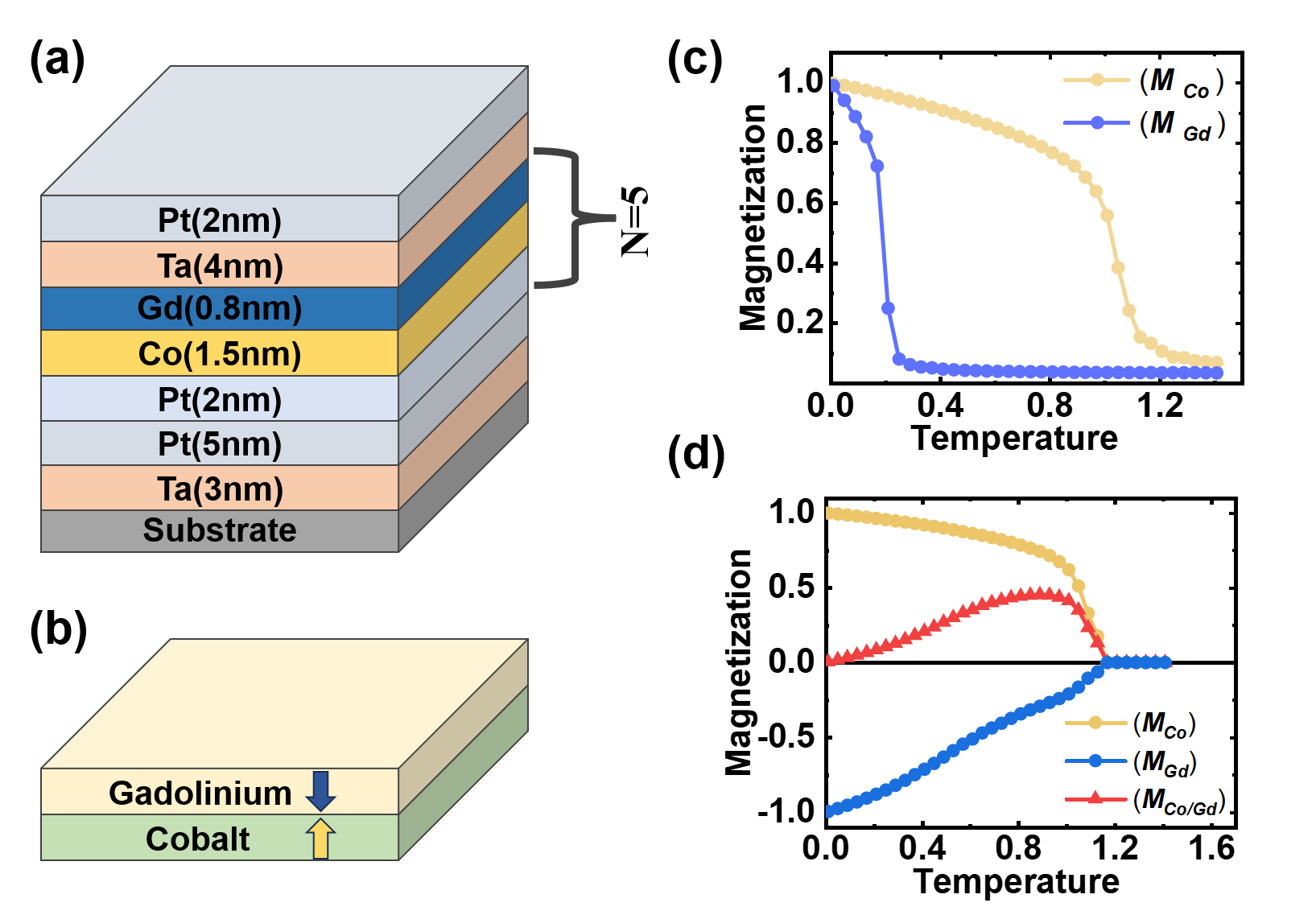}
 \caption{(a) Structural representation of sample FS, where N denotes one stacking unit of Pt/Co/Gd, repeated five times, (b) schematic illustration of spin alignment of Co and Gd in a SFiM, (c) simulated magnetization of the individual Co and Gd layers, highlighting their temperature dependence of magnetization (in units of the hopping amplitude), (d) simulated layer-resolved magnetization profiles of Co and Gd, after stacking as Co/Gd.}
\label{fig1}   
\end{figure*}

The operators $c_{i, l, \sigma}$ and $c_{i, l, \sigma}^\dagger$ represent the electronic annihilation and creation operators at site $i$ of layer $l$, respectively. The parameter $t^{(l)}$ is the nearest-neighbor hopping amplitude, while $J^{(l)}_{\rm H}$ characterizes the strength of Hund's coupling. The quantity $\lambda^{(l)}$ denotes the amplitude of the Rashba spin-orbit coupling. The vector $\boldsymbol{\tau}$ contains the three Pauli matrices as its components. The localized classical spin at site $i$ is represented by $\mathbf{S}_i^{(l)}$, and $\mathbf{s}_i^{(l)}$ denotes the electronic spin operator. The parameter $A^{(l)}_u$ controls PMA, and $h_z$ is the strength of the external magnetic field applied perpendicular to the triangular-lattice plane.

Assuming a unit lattice spacing, the vectors $\hat{\boldsymbol{\gamma}} \in \{ \mathbf{a}_1, \mathbf{a}_2, \mathbf{a}_3 \}$ correspond to the primitive vectors of the triangular Bravais lattice, with $\mathbf{a}_1 = (1,0)$, $\mathbf{a}_2 = (1/2,\sqrt{3}/2)$, and $\mathbf{a}_3 = (-1/2,\sqrt{3}/2)$. The hopping amplitudes $t^{(l)}$ are chosen to match the transition temperatures of bulk Co and Gd for layer $1$ and layer $2$, respectively. All the energy scales in the simulation are given in the unit of $t^{(1)}$.The term $H^{(l)}$ denotes the Hamiltonian for the individual layers, while $H_{\rm int}$ accounts for the interlayer interactions. The parameter $J_{\rm AF}$ specifies the antiferromagnetic exchange coupling between the two layers.
Assuming large Hund’s coupling limit\cite{Wu2022, Akosa2019, You2019, Wang2019} and 
following a procedure well-known for double-exchange models \cite{Kathyat2020, Kumar2006, Mukherjee2021}, we can write the effective Hamiltonain for individual layer as :  

\begin{align}   \label{eq:ESH}
H^{(l)}_{\rm eff} 
&= - D^{(l)}_{0}\sum_{\langle ij \rangle,\, \gamma} \, f^{(l)\gamma}_{ij}
   - \sum_{i} \left[ h_z\, S^{(l)z}_i + A^{(l)}_u \left(S^{(l)z}_i\right)^2 \right],
\nonumber \\[8pt]
\sqrt{2}\, f^{(l)\gamma}_{ij}
&= \Big[
\big(t^{(l)}\big)^{2} \left( 1 + {\bf S}^{(l)}_i \cdot {\bf S}^{(l)}_j \right)
+ 2\, t^{(l)} \lambda^{(l)}\, 
\hat{\boldsymbol{\gamma}}'\!\cdot\!\left( {\bf S}^{(l)}_i \times {\bf S}^{(l)}_j \right)
\nonumber \\[2pt]
&\quad
+ \big(\lambda^{(l)}\big)^{2} 
\left( 1 - {\bf S}^{(l)}_i \cdot {\bf S}^{(l)}_j
+ 2(\hat{\boldsymbol{\gamma}}' \cdot {\bf S}^{(l)}_i)\,
   (\hat{\boldsymbol{\gamma}}' \cdot {\bf S}^{(l)}_j) \right)
\Big]^{1/2}
\end{align}

In the above expressions, $f^{(l)\gamma}_{ij}$ denotes the modulus of the projected hopping amplitude, which depends on the relative orientations of the local moments ${\bf S}_i^{(l)}$ and ${\bf S}_j^{(l)}$. Earlier studies have shown that using constant coupling parameters yields an accurate description of the ground-state phases of $H^{(l)}_{\rm eff}$~\cite{Kathyat2020}. Accordingly, we set  $D_0^{(l)} = t^{(l)}/\sqrt{2}$ throughout our analysis. With this choice, the effective Hamiltonian becomes
\begin{equation}
H_{\rm eff} = \sum_{l=1}^{2} H^{(l)}_{\rm eff} + H_{\rm int}
\label{Hameff}
\end{equation}

We investigate Eq.~(\ref{Hameff}) using classical Monte Carlo simulations implemented via a standard Markov-chain Metropolis algorithm, further computational details are provided in Supplementary Note 3\cite{SM}.

\section{Results and discussion}
 The primary sample FS, is a five-repeat Pt/Co/Gd multilayer designed to host skyrmions by enhancing dipolar coupling and tuning the effective magnetic anisotropy. A schematic of this multilayer structure is shown in FIG. \ref{fig1}(a). As reference structures, two reference samples ($\rm{FR_1}$ and $\rm{FR_2}$) were fabricated, each consisting of a single Pt/Co/Gd trilayer, but with inverted capping sequences to isolate the influence of the top interface in sample FS. In $\rm{FR_1}$, Pt directly caps the magnetic layer, whereas in $\rm{FR_2}$, Ta is placed on top of Co/Gd followed by a thin Pt layer. These reference stacks help to disentangle the effects of multilayer repetition and capping-layer sequence on magnetic anisotropy and interfacial properties.

XRR patterns of sample FS display pronounced Bragg peaks and well-defined Kiessig fringes (FIG. S2, \cite{SM}), confirming the successful fabrication of periodic multilayer stacks and precise control over layer thicknesses. Single-trilayer reference samples $\rm{FR_1}$ and $\rm{FR_2}$ show the expected Kiessig fringes (FIG. S1, \cite{SM}), consistent with their simpler stack structure. This high degree of structural coherence is critical for maintaining consistent magnetic behaviour throughout the multilayer stack, which is essential for enabling the formation of nanoscale spin textures such as skyrmions. Further details about the XRR measurements are provided in the Supplementary Note 1 \cite{SM}.
Importantly, the Co/Gd bilayers introduce an additional degree of magnetic complexity due to their antiferromagnetic exchange coupling.  The net magnetization, $M_{\rm net}$, of the multilayer can be expressed as the vector sum of the Co and Gd sub-lattice magnetization, where the $M_{\rm Gd}$ points opposite to $M_{\rm Co}$. This coupling arises from the indirect antiferromagnetic exchange between the localized $4f$ moments of Gd and the itinerant $3d$ moments of Co, mediated through the hybridized $5d$ conduction electrons of Gd. This is mediated by the conduction electrons, resulting in a ferrimagnetic configuration of the multilayers (illustrated in FIG. \ref{fig1}(b)). In the present multilayer system, where a Gd layer deposited on the top of Co, the magnetic behavior differs markedly from that of bulk Gd or homogeneous Co-Gd alloys. Bulk Gd exhibits a Curie temperature of about 293 K, whereas Co retains ferromagnetic order up to approximately 1388 K. In order to resolve the two ordering scales in our simulations, we introduced a controlled separation of the Curie temperatures by tuning the tight-binding hopping amplitudes ($t^{(l)}$) of the Co and Gd sublayers. Since the hopping amplitude directly sets the electronic bandwidth, and hence the strength of the effective exchange interaction, increasing $t^{(l)}$ enhances the Curie temperature, while reducing $t^{(l)}$ suppresses it. In our calculations we chose hopping amplitudes of Co and Gd as 1.0 and 0.2, which allows us to set the transition temperatures approximately 1.0 and 0.2 (in the units of hopping amplitude), respectively (shown in FIG. \ref{fig1}(c)). This yields a ratio of $\approx$ 5, consistent with the experimental ratio of the Curie temperatures of Co and Gd. Consequently, as the temperature increases, the magnetization of the Co sublayer dominates while that of Gd gradually diminishes because of the lower Curie temperature than Co. Notably, the Gd magnetization does not collapse abruptly near its intrinsic Curie temperature ($\approx$0.2) but remains finite up to nearly the Co ordering temperature, owing to a proximity-induced stabilization from the ordered Co spins that suppress thermal fluctuations in Gd. FIG.\ref{fig1} (d) showing the interfacial proximity effect, where magnetic polarization extends across the Co/Gd interface due to strong antiferromagnetic exchange coupling.
 \begin{figure*}
\includegraphics[width=2.0\columnwidth,angle=0,clip=true]{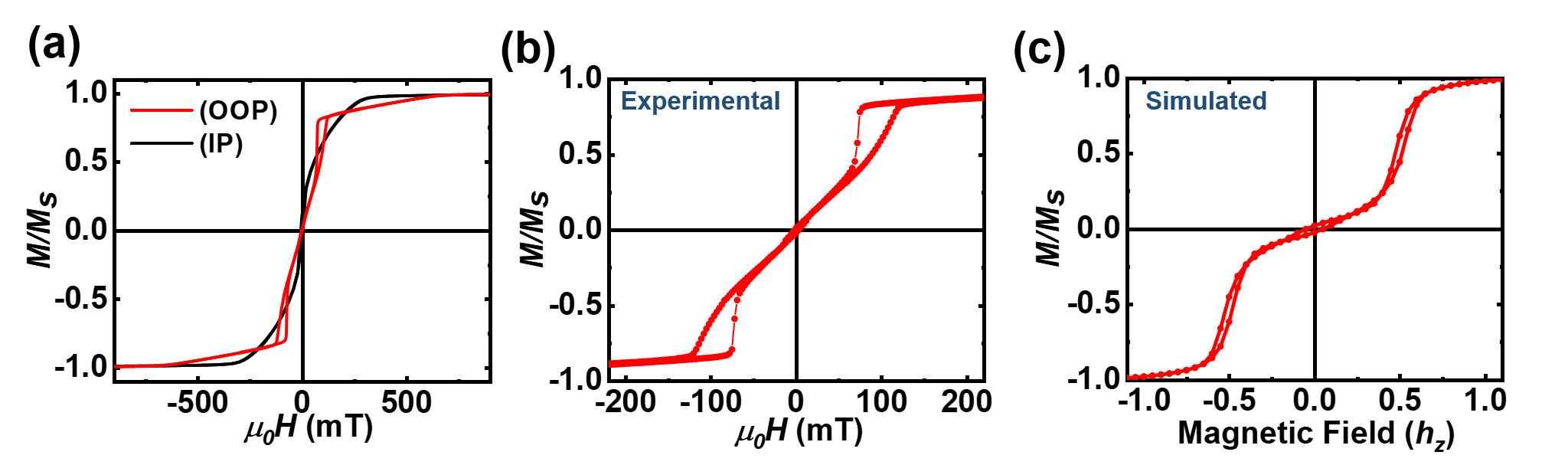}
 \caption{(a) Room temperature hysteresis loops for both OOP and IP magnetic fields for sample FS, (b) zoomed in image of OOP loop shown in (a) and, (c) hysteresis loop obtained from the simulation at effective temperature $T = 0.21$ approximating the room temperature.}
\label{fig2}   
\end{figure*}
\subsection{Room temperature hysteresis}
Further, the room-temperature hysteresis loops of sample FS reveal distinct anisotropic behaviour under OOP and IP magnetic fields [FIG. \ref{fig2}(a)]. The OOP loop exhibits a wasp-waisted shape, with a minor loop closing at $\mu_0H \approx 135$ mT due to the irreversible annihilation of skyrmionic textures stabilized by interfacial DMI and dipolar interactions shown in FIG. \ref{fig2}(b). Beyond this field, the magnetization gradually cants, requiring a much larger field ($\mu_0H \approx 700$ mT) to reach full OOP saturation, whereas the IP loop saturates at $\approx$ 500 mT , slightly preceding the OOP loop shown in [FIG. \ref{fig2}(a)]. This behavior can be understood by comparing the magnetic response of the reference samples. For the single-trilayer references, $\rm{FR_1}$ and $\rm{FR_2}$, hysteresis loops measured along OOP and IP directions reveal the influence of the capping layer on interfacial anisotropy. $\rm{FR_1}$ exhibits predominantly PMA, whereas $\rm{FR_2}$ shows in-plane anisotropy (IPA), highlighting how the topmost layer affects the effective anisotropy of the sample stack FS (shown in FIG. S3, with details in Supplementary Note 2 \cite{SM}). In sample FS, the multilayer design, consisting of five repetitions of Pt/Co/Gd, combined with a Ta capping layer on the topmost Gd, reduces perpendicular anisotropy. This arrangement lowers the effective anisotropy and places the system near to SRT, giving rise to the slanted feature in the OOP hysteresis loop, corresponding to a low-field skyrmion stage followed by gradual canting toward saturation.

In simulation, at temperatures close to the Curie point of Gd, the hysteresis acquires a three-loop structure (See FIG. \ref{fig2}(c)). The central loop originates from the reversal of the weakened Gd sublayer while the Co layer remains pinned, and this loop closes once the Gd magnetization has fully switched. At higher fields, the competition between the Zeeman energy and the antiferromagnetic interfacial exchange produces additional side loops, reflecting metastable configurations of the two sublayers. Ultimately, at sufficiently large fields, the Zeeman energy overcomes the interfacial exchange coupling and forces both Co and Gd moments into parallel alignment with the external field, closing the outer loops. 

\subsection{Temperature dependent net magnetization}

\begin{figure}[h!]
\includegraphics[width=1.0\columnwidth,angle=0,clip=true]{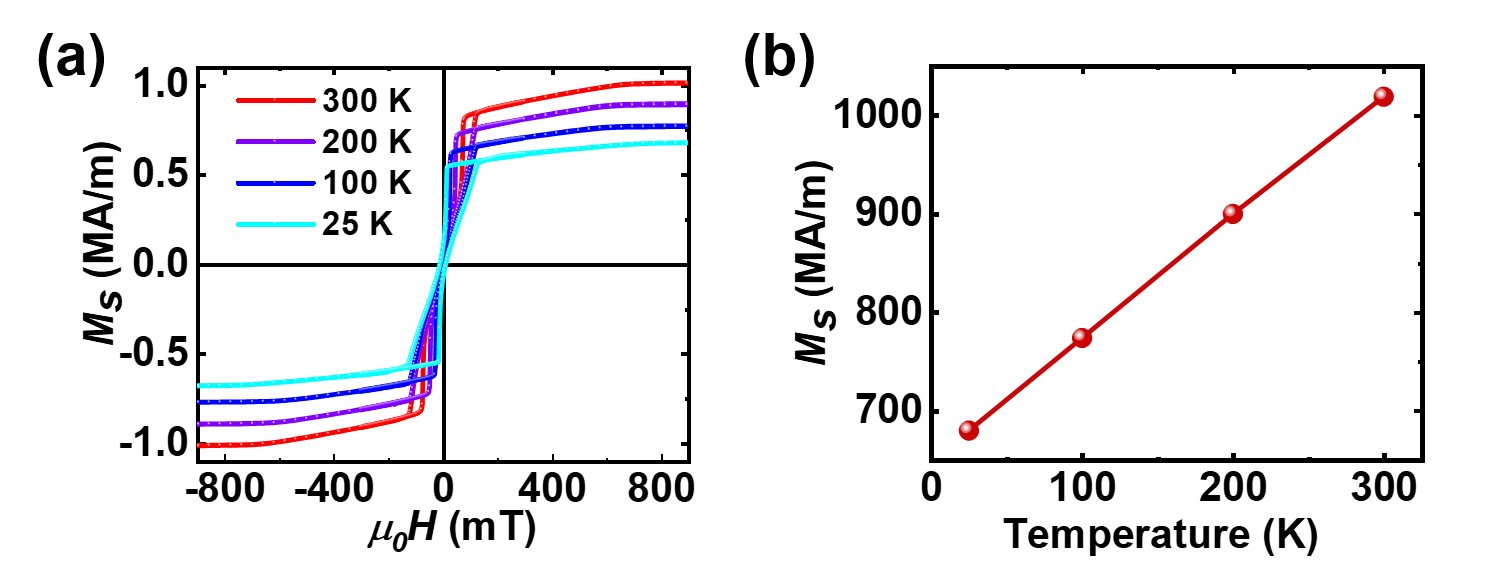}
 \caption{(a) Magnetic hysteresis loops recorded at 25 K, 100 K, 200 K, and 300 K, showing the temperature dependent evolution of magnetization for sample FS, (b) extracted saturation magnetization ($M_s$) plotted as a function of temperature, illustrating the thermal variation of the magnetic moment.}
\label{fig3}   
\end{figure}

Guided by these insights from simulations, we next examine the temperature-dependent magnetic behavior of sample FS. The corresponding hysteresis loops measured at 25 K, 100 K, 200 K, and 300 K are shown in FIG. \ref{fig3}(a), while FIG. \ref{fig3}(b) presents the extracted saturation magnetization from these loops and a plot of saturation magnetization (${M_s}$) versus temperature. Unlike conventional ferromagnetic multilayers, where net magnetization typically decreases with increasing temperature due to thermal disordering, the net magnetization of these SFiM multilayers increases with temperature, rising from 671 kA/m at 25 K to 1005 kA/m at 300 K.

 \begin{figure*}
\includegraphics[width=1.5 \columnwidth,angle=0,clip=true]{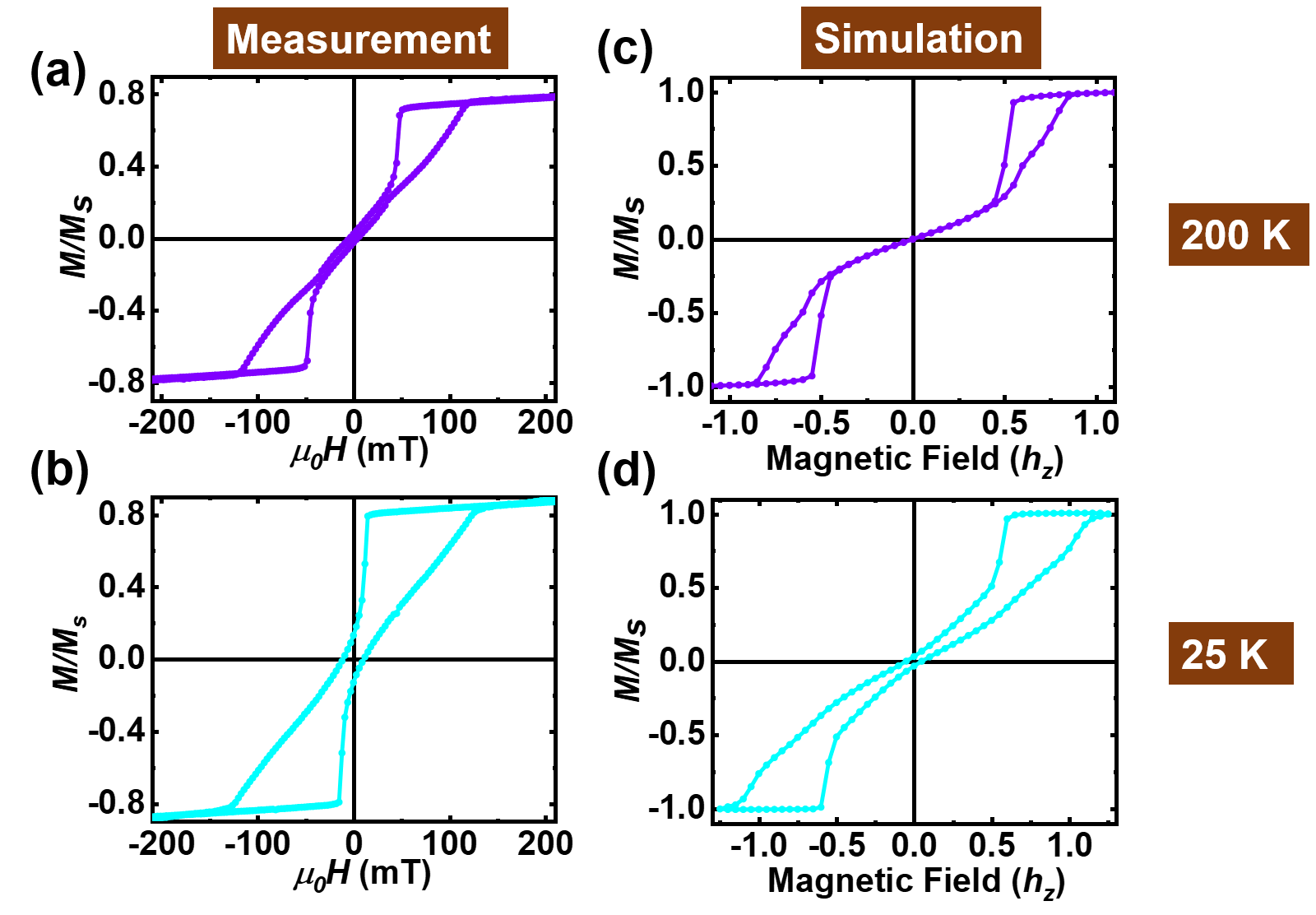}
 \caption{Experimental hysteresis loops of sample FS measured at (a) 200 K, and (b) 25 K, simulated hysteresis loop at effective temperature, (c) $T=0.09$ approximating to 200 K, (d) $T=0.01$ approximating to 25 K.}
\label{fig4}   
\end{figure*}

This unusual behavior is a hallmark of SFiM, arising from the ferrimagnetic coupling between the Co and Gd sublattices. At low temperatures, the strong Gd sublattice partially compensates more of Co moments, resulting in a lower net magnetization. As the temperature rises, the Gd moments decrease more rapidly due to their localized nature and lower thermal stability, causing the net magnetization to gradually increase (also evident in FIG. \ref{fig1}(c)). No compensation point is observed within the measured range, indicating that the Co sublattice dominates across the entire temperature window.
The thermal evolution of the hysteresis loops provides further insight into the magnetic behavior and is summarized in FIG. \ref{fig4}. Two notable trends emerge with decreasing temperature. First, the difference between the OOP and IP saturation fields gradually narrows with decreasing temperature: at room temperature, the OOP loop saturates at $\approx$ 700 mT and the IP loop at $\approx$ 500 mT, while the corresponding low-temperature loops are shown in FIG. S4 of the Supplementary Note 2 \cite{SM}. This reflects an increase in effective perpendicular anisotropy at lower temperatures, making the OOP axis energetically favored and easier to saturate than the in-plane direction (see Table 1, Supplementary Note 2 \cite{SM}). Although canting and dipolar interactions persist at low temperature, their influence is weaker, so the OOP orientation dominates the magnetization behavior. Second, the coercivity of the OOP minor (skyrmionic) loop increases at lower temperatures due to enhanced perpendicular anisotropy, primarily because the Gd sublattice exhibits a larger magnetic moment, reinforcing the out-of-plane alignment of the net magnetization. This increase in coercivity is directly evident in the experimental hysteresis loops measured at 200 K and 25 K [FIG. \ref{fig4}((a)-(b))], where the low-temperature loop exhibits a noticeably larger coercive field. This trend is reproduced in simulations, where the calculated hysteresis loops at equivalent temperatures(in terms of hopping amplitude) $T=0.09$ and $T=0.01$ [FIG. \ref{fig4}((c)-(d))] capture the experimentally observed increase in coercivity with decreasing temperature.\\
As discussed earlier, at low temperatures, the Gd moment increases and compensates a larger fraction of the Co magnetization, leading to a reduction in the net magnetization of the multilayer.
Although these opposing contributions largely cancel, a small net out-of-plane magnetization remains, arising from the uniform component of the spins along the easy axis, consistent with the low experimental $\rm{M_s}$ observed at 25 K. The Co sublayer, being stiffer due to its higher transition temperature, resists canting, whereas the Gd sublayer tilts slightly to minimize the interlayer exchange energy. This imperfect compensation of the uniform component gives rise to a finite macroscopic remanence, shown in FIG. \ref{fig4}(d). Details of the resulting spin textures at low temperature are discussed later (see FIG. \ref{fig8}). We note that the temperature dependence of magnetization obtained in our two layer model is in qualitative agreement with the experimental data. Furthermore, the experimentally reported M-H hysteresis behavior in different temperature regimes are also very well captured within the two layer model. These features justify our choice of minimal model for describing the magnetic properties of the SFiM.
\\

\subsection{Magnetic domain imaging at room temperature}
The constricted hysteresis loop in sample FS suggests the stabilization of nontrivial spin textures such as skyrmions. To directly visualize these textures and access their properties, we performed MFM measurements at RT. The measurements were carried out on a 5×5 $\mu m^2$ area with a lift height of 30 nm, enabling detection of the out-of-plane stray field distribution. This allowed identification of skyrmions, their size, distribution, and robustness under varying applied fields.
 \begin{figure}[h]
\includegraphics[width=0.94 \columnwidth,angle=0,clip=true]{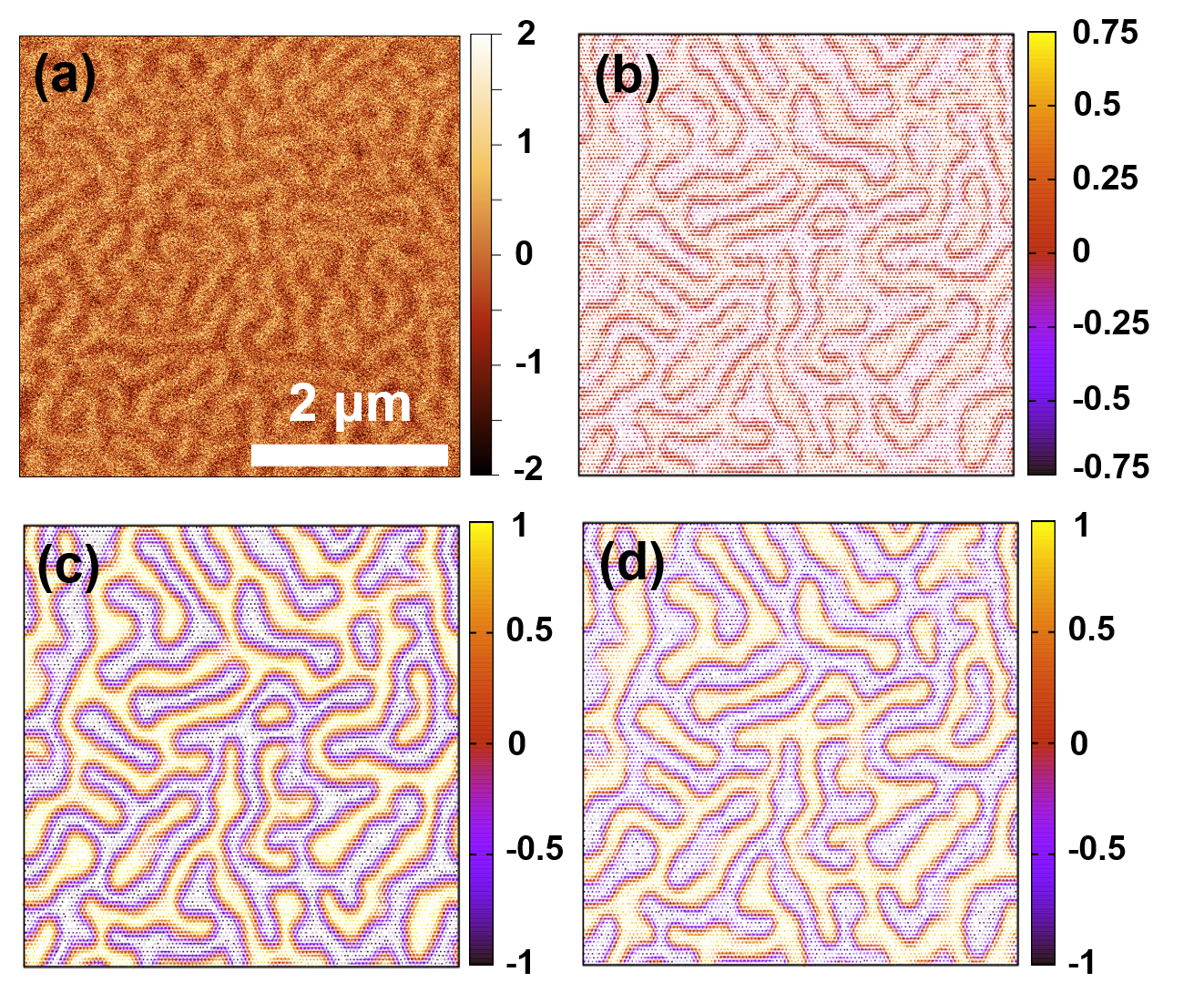}
 \caption{(a) Shows the experimental MFM image of sample FS in the demagnetized state at RT, while (b) presents the corresponding simulated demagnetized-state domain configuration at effective temperature $T=0.21$ , (c) and (d) represent simulated layer resolved domain structure in Co and Gd sublayers, respectively.}
\label{fig5}   
\end{figure}

The demagnetized state exhibits labyrinthine domains, with roughly equal areas showing upward and downward contrast, as shown in the experimental MFM image [FIG. \ref{fig5}(a)]. This domain pattern arises from the competition among PMA, exchange interactions, dipolar (demagnetizing) energy, and interfacial DMI. To further understand this domain state,  simulations were carried out at an effective temperature $T=0.21$, slightly above the ordering temperature of the Gd sublayer in our model. The parameters were chosen as $t^{(2)}/t^{(1)}=0.2$ and uniaxial anisotropy $A_u^{(1)}$=0.10. The spin–orbit coupling strength was set to $\lambda^{(l)}/t^{(l)}=0.40$, while the interlayer exchange coupling strength ${J_{\rm AF}}$ was fixed to 0.50. This parameter regime ensures a strong and nearly rigid ferromagnetic order in the Co sublayer, while the Gd sublayer remains comparatively soft and strongly temperature dependent. We have also made sure that the system size was large enough to suppress finite-size effects and can capture long-wavelength spin textures.

The simulated domain images of the demagnetized state [FIG. \ref{fig5}(b)] reproduce the labyrinthine pattern and reveal the domain structure in both the Co and Gd sublayers, highlighting the ferrimagnetic nature of the multilayer. In particular, the simulations show that the Co and Gd layers exhibit opposite magnetic contrast within the domains, directly reflecting their antiparallel spin alignment. This is explicitly illustrated in FIG. \ref{fig5}((c)–(d)), where panel (c) shows the domain image for the Co sublayer and panel (d) for the Gd sublayer.\\

 \begin{figure}[h!]
\includegraphics[width=0.96 \columnwidth,angle=0,clip=true]{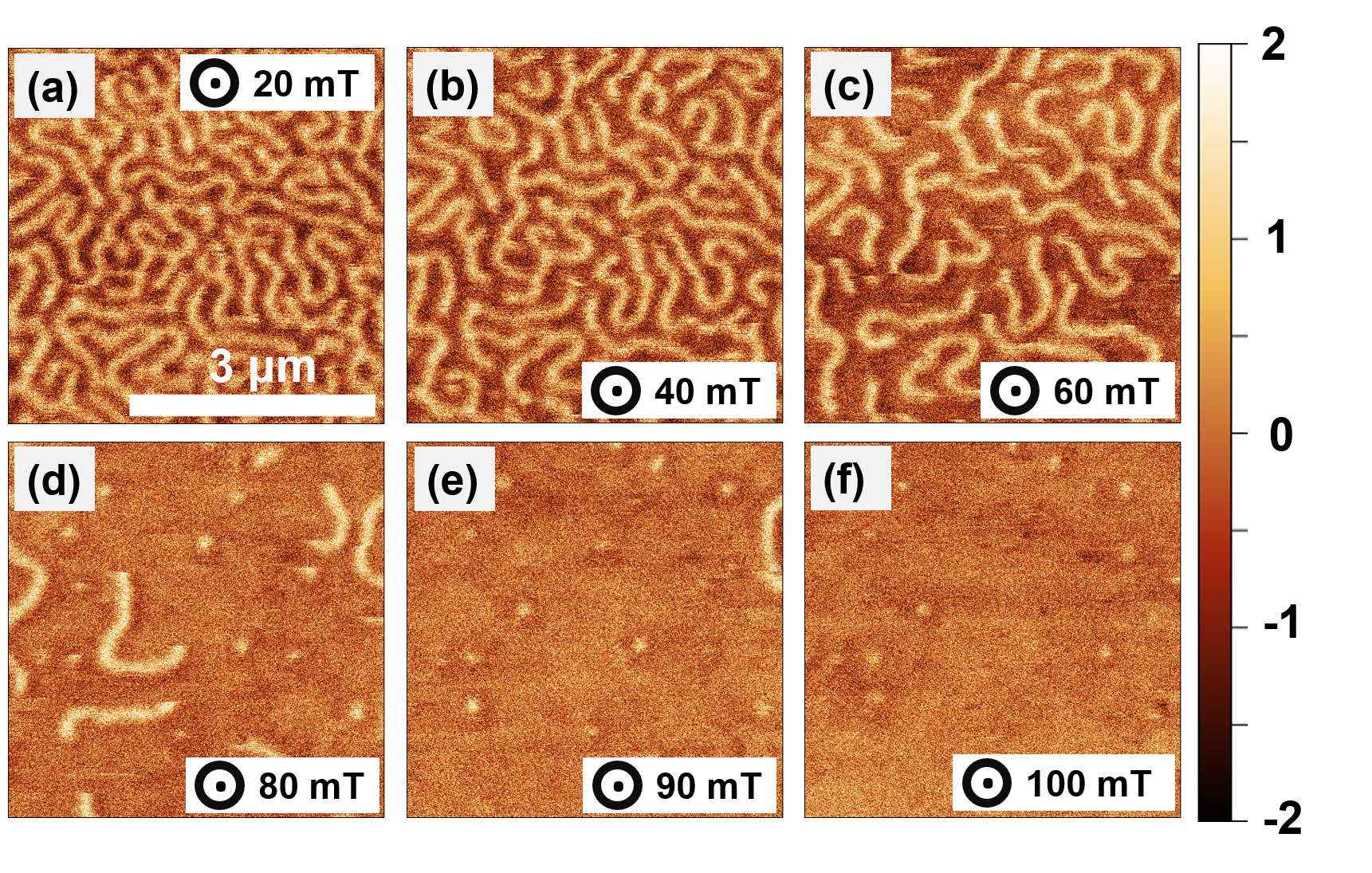}
 \caption{(a-f) Field evolution of skyrmionic textures in sample FS at RT in presence of out-of-plane magnetic fields, showing the breakup of labyrinthine domains into discrete skyrmions.}
\label{fig6}   
\end{figure}

Further, the evolution of skyrmionic textures under an applied OOP magnetic field was investigated using MFM and compared with simulated domain images. At low fields along the minor OOP loop, the labyrinthine domains begin to deform [FIG. \ref{fig6}(a-c)], and isolated skyrmions start to nucleate within the domain matrix [FIG. \ref{fig6}(d-f)]. As the field increases toward $\mu_{0}H \approx 70~\mathrm{mT}$, the labyrinth progressively breaks into discrete skyrmions, which appear as circular bright contrasts in the MFM images [FIG. \ref{fig6}(d-f)]. With increasing field, the skyrmions annihilate, and by $\mu_0 H \approx 135$~mT all textures vanish, consistent with the closure of the minor hysteresis loop. The skyrmion distribution is spatially non-uniform, reflecting the pinning landscape and local anisotropy variations. 

These isolated skyrmions, rather than a skyrmion lattice, appear when \(0 < \kappa < 1\), where the skyrmion stability parameter \(\kappa\) is defined as,
\begin{align}
    \kappa = \frac{\pi D}{4\sqrt{A K_{\mathrm{eff}}}}
    \label{Ham2}
\end{align}

$D$, $A$, and $K_\mathrm{eff}$ are the DMI, exchange stiffness, and effective magnetic anisotropy, respectively~\cite{Raju2021, soumyanarayanan2017}. Skyrmion lattices usually appear in bulk materials where $\kappa > 1$. A skyrmion lattice can appear in a thin film only when the DMI is sufficiently strong to overcome both exchange and anisotropy, making $\kappa \gtrsim 1$ possible. In most of the multilayers, magnetic anisotropy dominates over DMI, so only isolated metastable skyrmions are observed. \\
 \begin{figure}[h!]
\includegraphics[width=0.96 \columnwidth,angle=0,clip=true]{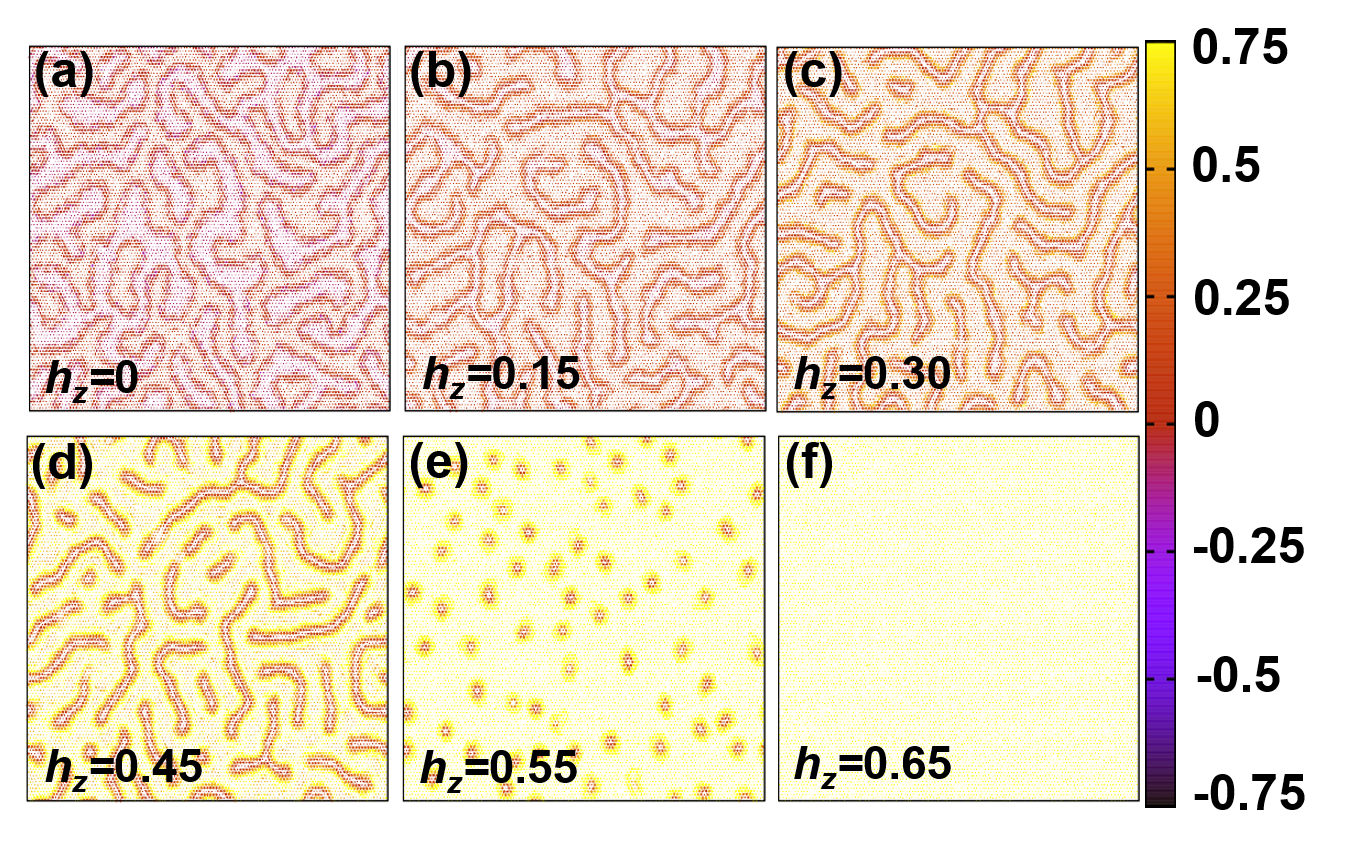}
 \caption{(a-f) Simulated domain configurations at effective temperature $T=0.21$ with the evolution of external magnetic field ($h_z$).}
\label{fig7}   
\end{figure}\\
Importantly, the simulated images [FIG. \ref{fig7}(a-f)] qualitatively reproduce the observed nucleation, evolution, and annihilation of skyrmions under the applied field. For these field-dependent simulations, the applied external magnetic field ($\rm{h_z}$) is indicated in the figures. The color bar corresponds to the z-component, and the arrows denote the planar spin components. In the model, the system was cooled in zero field, yielding a stable labyrinthine domain state near $T \approx 0.21 $, just above the ferromagnetic transition of bulk Gd. Upon increasing field, these labyrinths progressively turn into skyrmions, which were eventually annihilated at higher fields, producing a saturated out-of-plane state [FIG. \ref{fig7}(f)]. This field-driven transformation is consistent with the experimentally observed textures. 
 \begin{figure}[h!]
\includegraphics[width=0.96 \columnwidth,angle=0,clip=true]{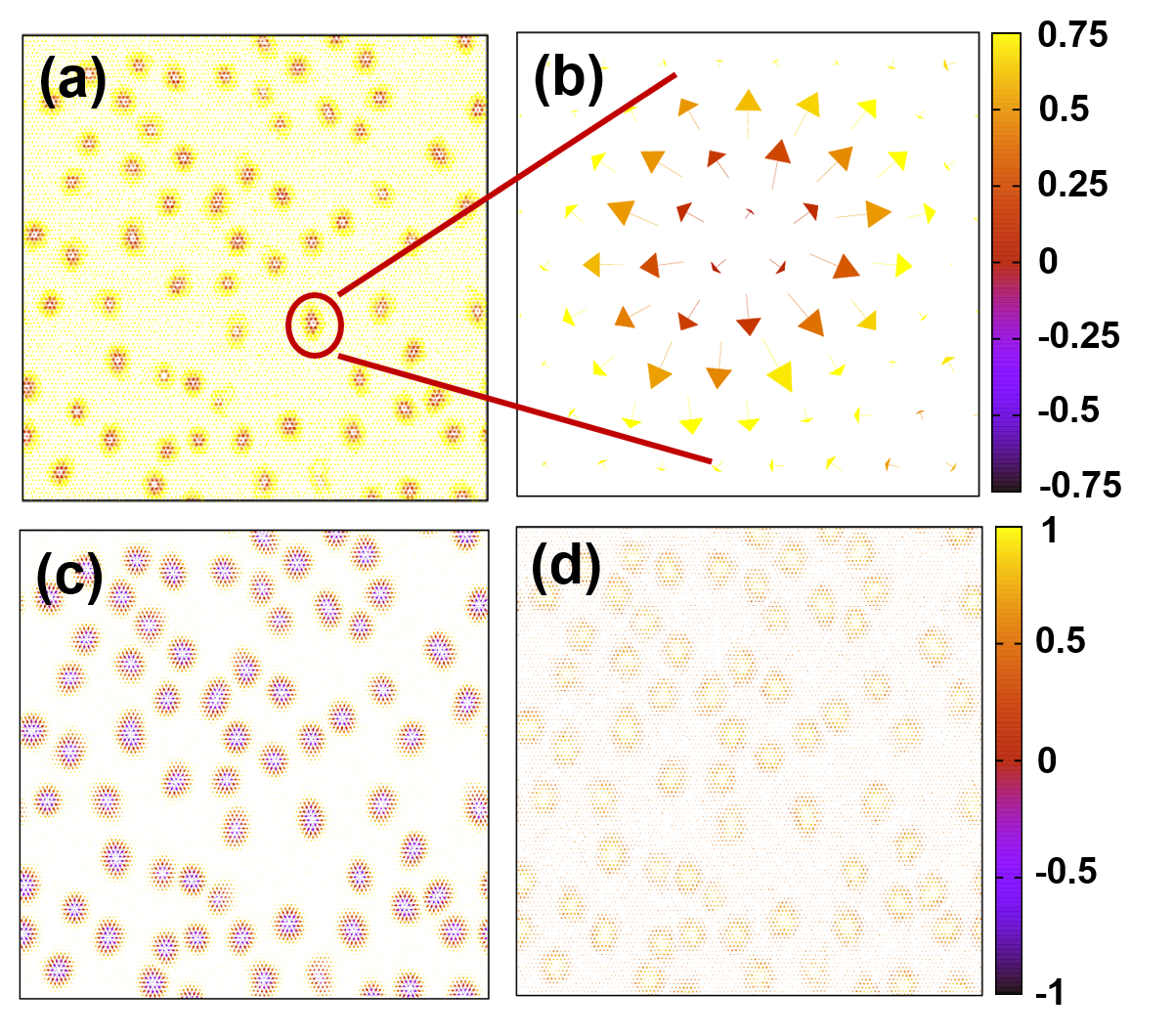}
 \caption{(a) Simulated skyrmion image at RT, (b) magnified view of  a single skyrmion, skyrmion image in the (c) Co layer, (d) Gd layer at effective temperature $T=0.21$.}
\label{fig8}   
\end{figure}Importantly, earlier work established that the domain width scales inversely with the strength of spin–orbit coupling \cite{Rana2023}. Hence, the experimental length scale can be captured in simulations by taking smaller SOC strength on larger lattices, without altering the essential physics. Additionally, the simulated state [FIG. \ref{fig8}(a)] with multiple skyrmions is further illustrated, where a zoomed view of one skyrmion [FIG. \ref{fig8}(b)] is presented to highlight its N\' eel-type spin configuration, with spins rotating radially from the core to the periphery. The corresponding layer-resolved textures for Co and Gd [FIG. \ref{fig8}(c-d)] further confirm that both sublayers host skyrmions with antiparallel alignment.\\

 \begin{figure}[h!]
\includegraphics[width=0.96 \columnwidth,angle=0,clip=true]{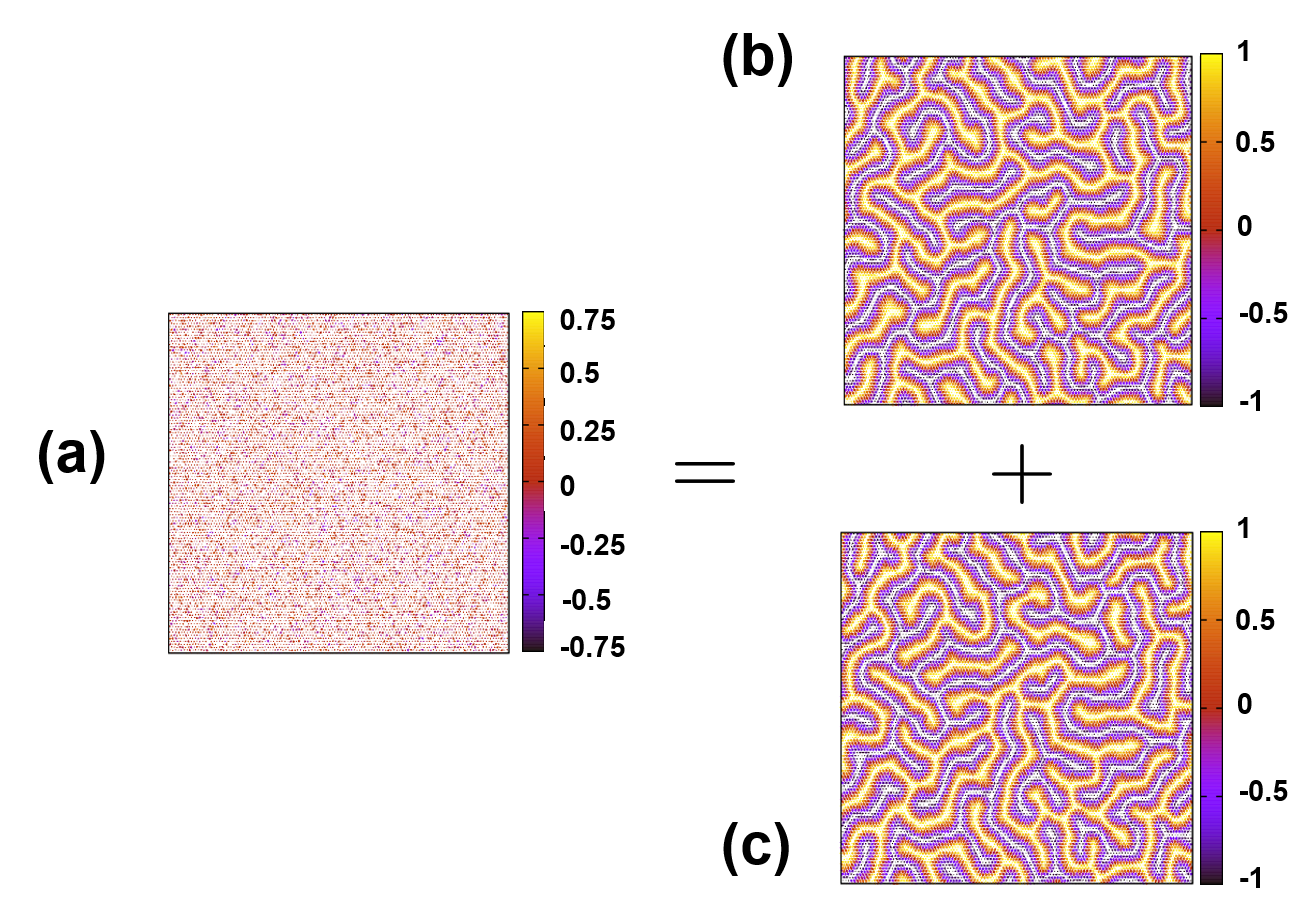}
 \caption{(a) Simulated demagnetized state domain image at effective tempertature $T=0.01$, layer-resolved domain structures of (b) Co and (c) Gd sublayers, at effective temperature $T$=0.01.}
\label{fig9}   
\end{figure}
Upon further cooling below the transition temperature of Gd, the domain contrast gradually diminishes and eventually disappears, reflecting the increasing magnetic moment of the Gd sublayer, which partially compensates the Co moments and reduces the net out-of-plane magnetization. In simulations at $T=0.01$ [FIG. \ref{fig9}(a)], the overall Co/Gd heterostructure in the demagnetized state exhibits no clear domain pattern, although separate inspection of the Co and Gd sublattices still reveals antiferromagnetically coupled domains [FIG. \ref{fig9}((b)-(c))]. Consequently, the labyrinthine and skyrmion-like textures become less pronounced, highlighting the delicate interplay of sublayer magnetization, interlayer coupling, and temperature in controlling the stability and visibility of magnetic textures in SFiM. This trend is rather unusual and interesting, since in most magnetic systems labyrinthine domain structures are stabilized at low temperatures and gradually disappear as thermal fluctuations increase. Our work shows that the separation of temperature scales between the sublayers and unequal response with the temperature opens a window where these magnetic textutres are stablilized.

\section{Conclusion}
In conclusion, our combined experimental and theoretical investigation provides a comprehensive study of the temperature-dependent magnetic behavior of Pt/Co/Gd multilayers. The multilayers exhibit anisotropy near the SRT and antiferromagnetic coupling between Co and Gd, forming a synthetic ferrimagnetic state. Magnetization and MFM measurements at room temperature reveal wasp-waisted hysteresis loops and isolated skyrmions stabilized by interfacial DMI and dipolar interaction. MFM images labyrinthine domains at zero field, which transform into isolated skyrmions of $\approx$ 70 nm radius under applied OOP fields. To interpret the observed field and temperature dependence magnetization and domain evolution, we developed a microscopic spin model incorporating spin–orbit coupling, uniaxial anisotropy, and interlayer exchange.  For flexibility and low computational cost, the system was represented as a two-layer triangular lattice, effectively capturing the Co–Gd antiferromagnetic coupling and DMI-driven energetics. Temperature-dependent studies highlight the pivotal role of the Gd sublayer: at low temperatures, enhanced Gd moments partially compensate Co magnetization, reducing domain contrast and increasing coercivity, whereas at higher temperatures, weakened Gd magnetization leads to stronger net moments and facilitates skyrmion nucleation. The simulations reproduce key experimental trends and reveal how the competition between thermal effects and interlayer exchange governs skyrmion stability. These results establish a robust framework for tuning magnetic interactions in SFiM, providing useful design principles for skyrmion-based spintronic devices. These magnetic systems provide a novel example of stabilization of topologically nontrivial textures at higher temperatures despite the ground states being topologically trivial. This puts in question the conventional theoretical approach for finding non-trivial electronic phases by first establishing the desired properties in the ground states and then finding ways to enhance the transition temperatures. Our work opens up possibilities to engineer phases with novel electronic and magnetic behavior, including topologically non-trivial phases, directly at room temperature without the existence of the corresponding ground states.

\begin{acknowledgments}
 S.B., B.S., S.M., B.O. thank the Department of Atomic Energy (DAE) of India, and DST-SERB (Grant No. CRG/2021/001245) for financial support. 
 S.B. and D.A. acknowledge the financial support by the Royal Society International Exchange scheme between Durham University, UK and NISER, Bhubaneswar, India, (No. IESnR1n201331). S.K. and S.D. acknowledge the use of high-performance Computing facility at IISER Mohali. 
\end{acknowledgments}

\bibliography{ref}
\end{document}